\documentclass[useAMS,usenatbib,usegraphicx]{mn2e}

\usepackage{graphicx}
\usepackage[latin1]{inputenc}
\usepackage{color}
\usepackage{times}
\usepackage{natbib}
\usepackage{setspace}
\newif\ifAMStwofonts
\AMStwofontstrue
\definecolor{red}{rgb}{1,0.,0.}

\newcommand{\morgana}{{\sc morgana}}
\newcommand{\munich}{WDL08}
\newcommand{\somer}{SHC08}
\newcommand{\kang}{KVB08}
\newcommand{\oiii}{{\rm OIII} }
\newcommand{\ha}{H\alpha }
\newcommand{\msun}{{\rm M}_\odot}

\def\lesssim{\lower.5ex\hbox{$\; \buildrel < \over \sim \;$}}
\def\gtrsim{\lower.5ex\hbox{$\; \buildrel > \over \sim \;$}}

\title[AGN in SAMs] {The dependence of AGN activity on stellar and
  halo mass in Semi-Analytic Models.}

\author[Fontanot et al.]{
  \parbox[t]{\textwidth}{ 
    Fabio Fontanot$^1$\thanks{Email: fontanot@oats.inaf.it},
    Anna Pasquali$^2$,
    Gabriella De Lucia$^1$, 
    Frank C. van den Bosch$^3,$
    Rachel S. Somerville$^{4,5}$,
    Xi Kang$^{6,2}$
    }
    \vspace*{6pt}\\
    $^1$ INAF-Osservatorio Astronomico di Trieste, Via Tiepolo 11, 
         I-34131 Trieste, Italy \\
    $^2$ MPIA Max-Planck-Institute f\"ur Astronomie, Koenigstuhl 17, 
         69117 Heidelberg, Germany\\
    $^3$ Department of Physics \& Astronomy, University of Utah, 
         115 South 1400 East, 201, Salt Lake City, UT 84112-0830, USA\\
    $^4$ Space Telescope Science Institute, 3700 San Martin Drive, 
         Baltimore, MD 21218, USA \\
    $^5$ Department of Physics and Astronomy, Johns Hopkins University, 
         Baltimore, MD 21218, USA \\
    $^6$ Purple Mountain Observatory, Chinese Academy of Sciences, 
         West Beijing Road 2, Nanjing 210008, China}

\begin{document}
\date{Accepted ... Received ...}

\maketitle

\begin{abstract} 
AGN feedback is believed to play an important role in shaping a
variety of observed galaxy properties, as well as the evolution of
their stellar masses and star formation rates. In particular, in the
current theoretical paradigm of galaxy formation, AGN feedback is
believed to play a crucial role in regulating star formation activity
in galaxies residing in relatively massive haloes, at low
redshift. Only in recent years, however, has detailed statistical
information on the dependence of galaxy activity on stellar mass
$M_\ast$, parent halo $M_{\rm DM}$ mass and hierarchy (i.e. centrals
or satellites) become available. In this paper, we compare the
fractions of galaxies belonging to different activity classes
(star-forming, AGN and radio active) with predictions from four
different and independently developed semi-analytical models. We adopt
empirical relations to convert physical properties into observables
($\ha$ emission lines, \oiii line strength and radio power). We
demonstrate that all models used in this study reproduce the observed
distributions of galaxies as a function of stellar mass and halo mass:
star forming galaxies and the strongest radio sources are
preferentially associated with low-mass and high-mass galaxies/haloes,
respectively. However, model predictions differ from observational
measurements in many respects. All models used in our study predict
that almost every $M_{\rm DM} > 10^{12} M_\odot$ dark matter halo
and/or $M_\ast>10^{11} M_\odot$ galaxy should host a bright radio
source, while only a small (few per cent) fraction of galaxies belong
to this class in the data. In addition, radio brightness is expected
to depend strongly on the mass of the parent halo mass in the models,
while strong and weak radio galaxies are found in similar environments
in data. Our results highlight that the distribution of AGN activity
as a function of stellar mass provides one of the most promising
discriminants between different gas accretion schemes.
\end{abstract}


\section{Introduction}\label{sec:intro}
In the last decade, a consensus has grown that there is a strong
correlation between the presence of active galactic nuclei (AGN)
powered by gas accretion onto Super Massive Black Holes (SMBHs), and
the properties of their host galaxies. From the observational
viewpoint, well defined correlations \citep{KormendyRichstone95,
  Magorrian98, FerrareseMerritt00, Gebhardt00, MarconiHunt03,
  HaringRix04} between the mass of SMBHs ($M_{\rm BH}$) and the
stellar mass, luminosity and velocity dispersion of the spheroidal
component of the host galaxy point towards a joint evolution of these
two components and a self regulated growth of the system. The details
of the relative growth of the two components are, however, still
uncertain (for a critical discussion, see
e.g.~\citealt{Peng06,Zheng09}).

From the theoretical viewpoint, AGNs have long been thought to
represent a transition phase in galaxy evolution, with no significant
influence on the evolution of the physical properties of galaxies as a
function of cosmic time. Nowadays, this view has changed, and AGNs are
viewed as a crucial ``ingredient'' in galaxy evolution. Very luminous
AGNs (i.e. quasars) release tremendous amounts of energy on very short
time-scales, eventually triggering powerful galactic winds, that can
deplete the cold gas reservoir of the host galaxies, and quench their
star formation activity \citep{DiMatteo05, MonacoFontanot05,
  PoundsPage06, GangulyBrotherton08}. In addition, in order to
reproduce the red colour and low levels of star formation of local
massive galaxies, a viable physical mechanism is required so as to
quench the predicted strong cooling flows in relatively massive haloes
at low redshift. Low luminosity radio AGNs provide an elegant solution
to this problem \citep{Bower06, Croton06, Sijacki07}: radio jets,
accelerated by the central engine, can efficiently transport energy
from central regions outward, offsetting cooling onto the central
regions \citep{BinneyTabor95, RuszkowskiBegelman02, KaiserBinney03}.

In galaxy formation models, high accretion events are usually
associated with galaxy mergers, that can destabilise large amounts of
cold gas in the colliding galaxies and drive it towards the centre of
the merger remnant. Observational data indicate a strong correlation
between mergers and enhanced star formation rates in ultra luminous
infrared galaxies \citep{SandersMirabel96, Barton00, Pasquali04,
  Woods06, Lin07, Li08b}. At the same time, theoretical studies have
pointed out that the rapid inflow of cold gas towards the centre may
feed the central SMBH and produce a luminous quasar phase
\citep{NegroponteWhite83, BarnesHernquist96, Monaco00, Granato01,
  Cattaneo05}. The feedback arising from this process is thermal,
i.e. originating from the coupling of a fraction of the QSO bolometric
luminosity with the interstellar medium. In addition, simulations show
that, in the case of a merger between two almost equal-mass galaxies,
the remnant morphology is spheroidal (i.e. dominated by random orbits,
\citealt{Barnes92}, \citealt{WalkerMihosHernquist96}). Therefore, if
the bulk of SMBH growth and the formation of the galactic spheroids
are both linked to major mergers, this scenario provides an elegant
explanation for the strong correlations observed between these two
galactic components \citep{KauffmannHaehnelt00,
  VolonteriHaardtMadau03, Hopkins05b, Hopkins06g}.

Another form of feedback, associated with radio activity, is believed
to play an important role in quenching cooling flows at the centre of
massive clusters. The development of radio jets might be connected
with low accretion rates, analogously to what happens for X-ray
binaries (see e.g. \citealt{Jester05}). In this scenario, the dominant
mode for AGN feedback is the injection of mechanical energy into the
ICM, associated with the development of X-ray cavities inflated by the
relativistic jets \citep{Birzan04, McNamaraNulsen07}. The physical
state of the accretion flow, as well as the spin of the SMBHs are
thought to play a role in determining the accretion rate efficiency
(see e.g. \citealt{Fanidakis09}), but details about the physical
mechanisms leading to the loss of angular momentum of the gas, the
subsequent accretion of material onto the SMBH, and the conversion of
mechanical into thermal energy are still to be understood. 

The physical processes mentioned above span from the Mpc scale of
galaxy mergers to the sub-pc scale of accretion discs and their
dependence on physical properties of galaxies (i.e. gas fractions, gas
density, morphological type) are still uncertain. Therefore, a
semi-analytic approach that couples a statistical (or numerical)
description of the growth of the cosmic structure with simple (yet
physically and/or observationally motivated) prescriptions for the
various physical processes at play, provides an efficient and valid
tool to test specific physical assumptions and to efficiently explore
the parameter space defined by each model.

The recent analysis of the SDSS data carried out by \citet{Pasquali09}
has opened new interesting prospects for a detailed comparison with
predictions from theoretical models. This analysis has taken advantage
of the group catalogue defined by \citet{Yang07} applying a halo-based
group finder \citep{Yang05} to the New York University Value-Added
Galaxy Catalogue \citep{Blanton05}, based on SDSS-DR4
\citep{Adelman-McCarthy06}. The group finder provides an estimate of
the parent Dark Matter (DM) halo mass ($M_{\rm DM}$) of each galaxy
group/cluster, as well as of the ``hierarchy'' (i.e. the central or
satellite nature) for all galaxies. Stellar masses estimates
($M_\star$) are computed using the observed relation between the
stellar mass-to-light ratio and the galaxy colour \citep{Bell03}. The
same group catalogue has been used in a series of paper
\citep{Weinmann06a, vdBosch08, Kimm08, Weinmann09, Pasquali10} to
investigate the dependence of galaxy properties on stellar mass,
environment and hierarchy. In particular, \citet{Kimm08} have used the
same group catalogue to study the empirical dependence of star
formation activity on internal galaxy properties and large-scale
environment, and have compared observational measurements with results
from semi-analytic models. These studies have shown that most models
including some prescriptions for AGN feedback reproduce reasonably
well the observed dependence of the red/passive fraction as a function
of stellar mass and halo mass for central galaxies. The same models,
however, produce too large a fraction of red and passive satellite
galaxies.

\citet[P09 hereafter]{Pasquali09} have extended this analysis by
matching the group catalogue with the catalogue of emission line
fluxes by \citet{Kauffmann03c}, with the National Radio Astronomy
Observatory Very Large Array Sky Survey (NVSS; \citealt{Condon98}),
and the Faint Images of the Radio Sky at Twenty centimetres (FIRST;
\citealt{Becker95}). P09 defined four different ``activity''
classes. Star-forming galaxies (SFGs) were separated from AGN host
galaxies using line-flux ratios $[O_{\rm III}] \lambda 5007 / H\beta$
and $[N_{\rm II}] \lambda 6584 / H\alpha$ (the so-called BPT diagram,
\citealt{Baldwin81}). Radio sources detected at $1.4 {\rm GHz}$ were
split into sources with low (LRA) and high (HRA) radio activity. Using
this matched catalogue, P09 studied the environmental dependence of
galaxy activity, as well as its dependence on the galaxy stellar
mass. They showed, in particular, that SFGs are preferentially
low-mass galaxies living in relatively small haloes, optical AGN
activity is linked to intermediate mass galaxies and environments,
while radio galaxies mainly reside in massive cluster galaxies. Given
the strong stellar-halo mass correlation for central galaxies, it is
unclear whether the activity levels for these galaxies are more
strongly related to their stellar or halo mass. As for satellite
galaxies, P09 found that the dependence of their galaxy activity on
halo mass is about four times weaker than that on stellar mass,
i.e. activity in satellites is suppressed with respect to central
galaxies of similar mass.

In this work, we present a detailed comparison between the P09
observational data and predictions of four different semi-analytic
models. This paper is organised as follows: in sec.~\ref{sec:models}
we describe the different semi-analytic models we consider in this
study, while in sec.~\ref{sec:conversion} we discuss our strategy for
converting model predictions into observational quantities. In
sec.~\ref{sec:results}, we then compare these predictions with the
results presented in P09. Finally, we discuss our results and give our
conclusions in sec.~\ref{sec:final}.

\section{Models}\label{sec:models}

In this paper, we consider predictions from four independently
developed theoretical models for galaxy formation. In particular, we
consider the most recent implementations of four different
Semi-Analytic Models (SAMs, see e.g. \citealt{Baugh06} for a recent
review) for galaxy formation and evolution, in the concordance
$\Lambda$CDM cosmology: \morgana~\citep{Monaco07}, the \citet{Kang08a}
model (\kang~hereafter), the \citet{Somerville08} model
(\somer~hereafter), the \citet{Wang08} model (\munich~hereafter). In
these models, the evolution of the baryonic component is followed by
means of an approximate description of the physical processes at play
(such as gas cooling, star formation and feedback) and of their
interplay with gravitational processes (i.e. dynamical friction, tidal
stripping and two body mergers), linked to the assembly of the large
scale structure of the Universe. These ``recipes'' include a number of
parameters which are usually fixed by comparing model predictions with
a set of low-redshift observations. Despite their simplified approach,
SAMs have turned into a flexible and widely used tool to explore a
broad range of specific physical assumptions, as well as the interplay
between different physical processes.

We refer to the original papers for a detailed description of the
modelling adopted for the various physical processes considered (see
also \citealt{Fontanot09b} for a comparison between the \morgana,
\munich~and \somer~models). In the following, we will focus on the
modelling of gas accretion onto SMBHs and different modes of AGN
feedback. All models are based on an analytical or numerical
description for the redshift evolution of the mass and number density
of dark matter haloes (i.e merger trees\footnote{The \munich~and
  \kang~models use merger trees extracted from N-body simulations
  (\citealt{Wang08} and \citealt{JingSuto02} respectively);
  \morgana~uses the Lagrangian semi-analytic code {\sc pinocchio}
  \citep{Monaco02}, and \somer~use a method based on the Extended
  Press-Schechter formalism, described in
  \citet{SomervilleKolatt99}.}). The cosmological parameters adopted
are slightly different (see table~\ref{tab:cosmo}); we make no attempt
to correct the results, since we expect these small differences to
have negligible influence on model predictions \citep{Wang08}. In all
cases, the mass resolution is sufficient to resolve galaxies with
stellar mass larger than $10^9 \msun$. In the following, we
parameterise the Hubble constant as $H_0 = 100\, h\, {\rm km\, s^{-1}
  Mpc^{-1}}$.
\begin{table}
\begin{center}
  \begin{tabular}{ccccc}
    \hline
    Model & $\Omega_0$ & $\Omega_\Lambda$ & $h$ & $\sigma_8$ \\
    \hline
    \citet{Pasquali09} (P09)   & $0.238$ & $0.762$ & --- & --- \\
    \morgana             & $0.24$ & $0.76$ & $0.73$ & $0.80$ \\
    \citet{Kang05}       & $0.25$ & $0.75$ & $0.71$ & $0.75$ \\
    \citet{Somerville08} & $0.28$ & $0.72$ & $0.70$ & $0.817$ \\
    \citet{Wang08}       & $0.226$ & $0.774$ & $0.743$ & $0.722$ \\
    \hline
  \end{tabular}
  \caption{Cosmological parameters adopted by the semi-analytic
    realizations.}
  \label{tab:cosmo}
\end{center}
\end{table}

Gas inside DM haloes cools and condenses primarily via atomic cooling
and thermal Bremsstrahlung, and forms a rotationally supported
disc. Star formation is modelled using simple empirical
(Schmidt-Kennicutt-like) recipes; supernovae and stellar winds deposit
both thermal and kinetic energy into the cold gas and might re-heat or
expel part of this gas; galaxy mergers might trigger ``bursts'' of
star formation and play an important role in triggering AGN
activity. The details of the modelling of these processes differ among
the four SAMs we consider in this paper.

We recall that the \munich~and \kang~models explicitly follow the
evolution of DM substructures, until tidal truncation and stripping
reduce their mass below the resolution limit of the simulation (for
details, see \citealt{DeLucia07b} and \citealt{Kang05}
respectively). Satellite galaxies are then assigned a residual
survival time, which is roughly twice\footnote{A fudge factor of $\sim
  2$ accounts for recent findings that the classical formula
  systematically under-estimates merger times computed from numerical
  simulations \citep[e.g.][]{BoylanKolchin08,Jiang08}} the value
computed using the classical dynamical friction formula
\citep{BinneyTremaine87}. \somer~and \morgana~model the evolution of
DM substructures using analytic recipes based on the fitting formulae
proposed by \citet{BoylanKolchin08} and \citet{Taffoni03},
respectively. A detailed comparison between these different
formulations is given in \citet{DeLucia10}.

All models follow the growth of SMBHs at the centre of model galaxies,
and differentiate between the so-called ``bright-mode'' (or
``QSO-mode'') and ``radio-mode'' associated with the efficient
production of radio jets. In the following, we discuss in more detail
the modelling adopted for these physical processes.

\begin{itemize}

\item{{\bf \citet{Wang08}.} This model represents a generalisation of
  the \citet{DeLucia07b} model to the WMAP3 cosmology assumed in the
  \citet{Wang08} simulations. Predictions from the model presented in
  \citet{DeLucia07b} applied to the Millennium Simulation are publicly
  available\footnote{see
    http://www.mpa-garching.mpg.de/millennium/}. In this model, the
  ``bright-mode'' is triggered by galaxy-galaxy mergers following the
  approach originally proposed by \citet{KauffmannHaehnelt00}, and
  extended to minor mergers by \citet{Croton06}: during each merger
  (both minor and major) a fraction $f_{\rm BH}$ of the cold gas
  $M_{\rm cold}$ in the progenitors is assumed to be accreted on the
  SMBH at the centre of the merger remnant (the pre-existing SMBHs are
  assumed to merge instantaneously). The total accreted mass during
  the merger event is parameterised as:
  \begin{equation}\label{eq:kh00}
    \Delta m^{\rm C06}_{\rm QM} = f_{\rm BH} \, \zeta_m \,
    \frac{M_{\rm cold}}{1+\left( \frac{V_{\rm vir}}{280 {\rm km/s}} \right)^{-2}}
  \end{equation}
  where $\zeta_m$ represents the baryonic ratio between the progenitor
  galaxies (defined such as $\zeta_m<1$). According to the above
  equation, equal mass mergers result in stronger accretion rates than
  minor mergers between galaxies with similar amounts of cold
  gas. This is to be expected as numerical simulations show that major
  mergers are responsible for a stronger dynamical response and a more
  rapid infall of gas towards the centre. $f_{\rm BH}=0.03$ is a free
  parameter, chosen by requiring the model to fit the local relation
  between the BH mass and the mass of the hosting spheroid
  \citep{HaringRix04}.

  ``Radio-mode'' feedback is implemented as in \citet{Croton06} and is
  assumed to result from the accretion onto the central BH of hot gas
  from a static atmosphere. The accretion is assumed to be continuous
  and is described by the phenomenological recipe:
  \begin{equation}\label{eq:c06}
    \dot{m}^{\rm C06}_{\rm RM} = \kappa_{\rm C06} \left( \frac{M_{\rm BH}}{10^8
      M_\odot} \right) \left( \frac{f_{\rm hot}}{0.1} \right) \left(
    \frac{V_{\rm vir}}{200 {\rm km/s}} \right)^3
  \end{equation} 
  where $f_{\rm hot}$ is the fraction of the total halo mass in form
  of hot gas and $\kappa_{\rm C06} = 6.5 \times 10^{-6} M_\odot/{\rm
    yr}$. We note that in their original paper, \citet{Croton06}
  demonstrate that this formulation is consistent with expectations of
  theoretical models based on the analytical results of
  \citet{Bertschinger89} and on a Bondi-Hoyle model for BH
  feeding. The amount of energy injected by the AGN into the intra
  cluster medium (ICM) through this channel is:
  \begin{equation}
    L^{\rm C06}_{\rm RM} = \epsilon \, \dot{m}^{\rm C06}_{\rm RM} \,
    c^2
  \end{equation}
  where $\epsilon=0.1$ represents the conversion efficiency of rest
  mass to energy. The efficiency of gas heating by AGN is assumed to
  be independent of the halo mass.}

\item{{\bf \citet{Kang08a}.} The ``bright-mode'' is modelled as in
  \citet{KauffmannHaehnelt00} and is associated with major and minor
  mergers:
  \begin{equation}\label{eq:k06}
    \Delta m^{\rm K06}_{\rm QM} = g_{\rm BH} \frac{M_{\rm
        cold}}{1+(\frac{V_{\rm vir}}{280 {\rm km/s}})^{-2}}
  \end{equation}
  $g_{\rm BH}=0.03$ is a free parameter, again calibrated using the
  observed relation between the mass of the BH and the mass of hosting
  spheroid.

The ``radio-mode'' is modelled assuming that the energy injected from
the central SMBH is proportional to its Eddington Luminosity $L_{\rm
  edd}$ and that the heating efficiency scales as a power law of
$V_{\rm vir}$ \citep{Kang06}:
  \begin{equation}\label{eq:k08}
    L^{\rm K06}_{\rm RM} = \kappa_{\rm K06} \, L_{\rm edd} \, \left( \frac{V_{\rm vir}}{200
      {\rm km/s}} \right)^4
  \end{equation} 
  where $\kappa_{\rm K06} = 2 \times 10^{-5}$ is a free parameter,
  fixed by requiring the model to fit the bright end of the local
  galaxy luminosity function.

  Another important feature of this model is the treatment of hot gas
  associated to recently accreted satellites. The other models adopted
  in this study assume that the hot gas reservoir is instantaneously
  stripped when galaxies are accreted onto larger systems and become
  satellites. As argued by many authors (see e.g
  \citealt{Weinmann06a}), this assumption is probably too strong, and
  results in red satellite fractions significantly higher than
  observed. \kang~show that by assuming a constant stripping rate over
  a timescale of $\sim3$ Gyrs, the model is able to predict a fraction
  of blue satellites which is in qualitative agreement with
  observations (see also \citealt{Wang07}).}

\item{{\bf \citet{Somerville08}.} The model used in this study is
  discussed in detail in \somer, but uses the updated bulge formation
  model discussed in \citet{Hopkins09b}. Other minor changes include a
  new calibration of the galaxy formation parameters, appropriate for
  a WMAP5 cosmology. In this model, the ``bright-mode'' is associated
  with merger events, following the prescriptions proposed by
  \citet{Hopkins07a} and based on a suite of $N$-body simulations of
  galaxy-galaxy mergers \citep{Robertson06, Cox06}. For each merger
  event, the mass of the final spheroidal component $M_{\rm sph}$ is
  computed following \citet{Hopkins09a}, and the final mass of the
  SMBH is estimated using the simulation-calibrated relation
  \citep{Hopkins07a}:
  \begin{equation}
    \log(M_{\rm BH}/M_{\rm sph}) = -3.27+0.36 \, {\rm erf}[(f_{\rm gas}-0.4)/0.28]
  \end{equation} 
  where $f_{\rm gas}$ represents the ``effective'' gas fraction
  (defined as the sum of the cold gas masses in the merging galaxies,
  divided by their total baryonic masses). The accretion of gas onto
  the SMBH is modelled through the generalised functional form for QSO
  light curves proposed by \citet{Hopkins05b}: the SMBH grows at the
  Eddington rate, until it reaches a critical mass $M_{\rm BH,{\rm
      crit}}$. Further accretion is then reduced by the onset of a
  pressure driven outflow (which eventually halts the accretion) and
  this regime is described by a power-law decline. The critical mass
  is defined as:
  \begin{equation}
    M_{\rm BH,crit} = 1.07 \, f_{\rm BH,crit} \, (M_{\rm BH}/10^9 M_\odot)
  \end{equation}
  where $f_{\rm BH, crit} = 0.4$ is a free parameter. In this
  scenario, bright QSOs spend most of their lifetime in the power-law
  decline phase, but most of the BH mass is accreted at the Eddington
  rate. If the mass of the SMBH is already larger than the critical
  mass, no accretion is allowed on the central object. This might
  happen frequently at low redshift, where the most massive spheroidal
  galaxies have already assembled most of their stellar mass.

  The ``radio-mode'' feedback implementation is based on the
  assumption that low-accretion events are fuelled by Bondi-Hoyle
  accretion:
  \begin{equation}\label{eq:bondi}
    \dot{m}_{\rm bondi} = \pi \, (G M_{\rm BH}) \, \rho(r_{\rm A}) \, c_s^{-3}
  \end{equation}
  where $c_s$ represents the gas sound speed and $\rho(r_{\rm A})$ is
  the density of the hot gas at the accretion radius. By combining
  eq.~\ref{eq:bondi} with the isothermal cooling flow solution
  \citep{NulsenFabian00}, it is possible to rewrite the accretion rate
  as:
  \begin{equation}
    \dot{m}^{\rm S08}_{\rm RM} = \kappa_{\rm S08} \left(
    \frac{kT}{\Lambda(T,Z_h)} \right) \left( \frac{M_{\rm BH}}{10^8
      M_\odot} \right)
  \end{equation}
  where $T$ is the virial temperature of the hot gas, $k$ is the
  Boltzmann constant, and $\Lambda(T,Z_h)$ represents the cooling
  rate. $\kappa_{\rm S08} = 3.5 \times 10^{-3}$ is a free
  parameter. The energy that is effectively injected into the ICM (the
  heating rate) is obtained by converting the accreted matter into
  energy:
  \begin{equation}
  L^{\rm S08}_{\rm RM} = \epsilon \,\kappa_{\rm heat} \, \dot{m}^{\rm
    S08}_{\rm RM} \, c^2
  \end{equation}
  In the standard model, $\kappa_{\rm heat}=1.0$ and $\epsilon=0.1$.}
 
\item{{\bf \morgana.} In this model, the growth of SMBHs is modelled
  following the prescriptions by \citet{Umemura00} and
  \citet{Granato04} (see \citealt{Fontanot06} for more details on the
  actual implementations of these prescriptions in \morgana). The
  model follows the evolution of a cold gas reservoir around the SMBH,
  which is assumed to form through the same physical processes that
  drive the formation of the bulge. The formation rate of the
  reservoir $\dot{m}_{\rm resv}$ is then assumed to be proportional to
  the star formation rate in the bulge $\dot{m}_{\star,B}$:
  \begin{equation}
    \dot{m}_{\rm resv} = f_{\rm resv} \, \dot{m}_{\star,B} \, \left(
    \frac{\dot{m}_{\star,B}}{100 M_\odot} \right)^{\alpha_{\rm resv}-1}
  \end{equation}
  where $f_{\rm resv} = 3 \times 10^{-2}$ and $\alpha_{\rm resv}=1$
  are free parameters\footnote{Note that with this parameter choice,
    the equation simplifies to: $\dot{m}_{\rm resv} = f_{\rm resv} \,
    \dot{m}_{\star,B}$.}, fixed by requiring the model to reproduce
  the $0<z<5$ evolution of the QSO luminosity function
  \citep{Fontanot06}. The cold gas mass in the reservoir $M_{\rm
    resv}$ can be accreted onto the SMBHs, once it loses its residual
  angular momentum, through turbulence, magnetic field or radiation
  drag. The BH accretion rate is assumed to be determined by the
  viscosity of the accretion disc and is written as \citep{Granato04}:
  \begin{equation}\label{eq:morgana}
    \dot{m}^{\rm M07}_{\rm BH} = \kappa_{\rm visc} \left( \frac{\sigma_B^3}{G}
    \right) \left( \frac{M_{\rm resv}}{M_{\rm BH}} \right)^{3/2} \left(
    1+\frac{M_{\rm BH}}{M_{\rm resv}} \right)^{1/2}
  \end{equation}
  where $\sigma_B$ is the velocity dispersion of the bulge, and
  $\kappa_{\rm visc} = 10^{-3}$ is chosen following the theoretical
  arguments by \citet{Granato01}. The accretion rate is limited at the
  Eddington rate. In this model, it is the rate of accretion that
  determines the `nature' of the feedback: the ``bright-mode'' is
  associated with accretion rates larger than 1 per cent of the
  Eddington limit, while the ``radio-mode'' is sustained by lower
  accretion rates. For the radio-mode, \morgana~assumes a bolometric
  luminosity:
  \begin{equation}
  L^{\rm M07}_{\rm RM} = \epsilon \, \kappa_{\rm M07} \, \dot{m}^{\rm
    M07}_{\rm BH} \, c^2
  \end{equation}
  The efficiency ($\kappa_{\rm M07}$) at which the emitted energy
  heats the hot halo component is assumed to scale with the circular
  velocity of the host halo and, in the standard implementation, is
  normalised to the value at $10^3 {\rm km/s}$
  \begin{equation}\label{eq:finj}
    \kappa_{\rm M07} = \left( \frac{V_{\rm vir}}{1000 {\rm km/s}} \right)^3
  \end{equation}
  For consistency with the other models, in the following we neglect
  this virial velocity dependence (i.e. we set $\kappa_{\rm
    M07}=1.0$). We have verified, however, that its inclusion does not
  change significantly the results discussed below.

  For high accretion rates, the model assumes that only 10 per cent of
  the bolometric luminosity is available to heat the halo gas
  (considering that about 10 per cent of quasars are radio loud). We
  note that, at variance with the other models used in this study, in
  \morgana~the gas accreted during the ``radio-mode'' is cold. In
  addition, gas accretion is always associated with some star
  formation, responsible for the final loss of angular momentum of the
  accreted material. As a result, massive central galaxies in this
  model are too blue (because they form too many stars) with respect
  to the observational determinations.}

\end{itemize}

In this paper, we carry out a comparison between predictions from the
4 models introduced above and the observational measurements presented
in P09. We stress that none of the models used has been tuned to
reproduce the observational data considered. Therefore, model results
discussed below should be considered as genuine model predictions.

This section illustrates that the implementations of SMBH growth and
AGN feedback adopted in the models considered in this study, differ in
a number of details. We note, in particular, that minor mergers do
trigger accretion onto BHs in three models used in this study
(\morgana, \somer~and \munich), satellite-satellite merger are
considered in \kang~and \munich, while only \morgana~considers also
accretion of gas triggered by disk instabilities\footnote{We note that
  disk instabilities represent the main channel for SMBH growth in the
  \citet{Bower06} model, which is not considered in this study.}. In
addition, AGN driven winds are explicitly included only in
\morgana~and \somer. The \somer~model includes a more sophisticated
treatment for the evolution of the AGN during the ``bright'' phase,
explicitly including a description for the AGN light curve.

The models also share a number of similarities. ``Bright-mode''
feedback is tightly linked to galaxy-galaxy mergers in \munich,
\kang~and \somer. \morgana~adopts a different scheme, but also for
this model the high accretion rates needed in the ``bright-mode''
regime are often associated with galaxy-galaxy mergers. In addition,
three of the four models used in this study (\munich, \kang~and
\somer) assume (either directly or indirectly as in the \somer~model)
a strong relation between the mass of the parent DM halo and the
strength of the ``radio-mode'' feedback.

The variety of assumptions made to model these processes reflect the
fact that they are still not well understood, and that almost all
details of the physical mechanisms acting on sub-pc scale (i.e.  the
final stages of angular momentum loss, the properties of the inner
accretion disc, the AGN light curves, and the collimation of the radio
jet) are still largely unconstrained.

\section{Converting model predictions into observed quantities}
\label{sec:conversion}

In order to compare model predictions with the P09 results, we need to
convert the predicted accretion and star formation rates into
`observables'. These include, in particular, the luminosity of the
$\ha$ and \oiii lines, and the radio power at 1.4 {\rm GHz} $P_{\rm
  1.4GHz}$ used for the estimate of the star formation, AGN and radio
activity, respectively. All model predictions presented below are
constructed by considering a cosmological box centred at $z=0.1$, and
converting predicted intrinsic luminosities into apparent fluxes.

\subsection{\oiii line luminosity}\label{lo3}
\begin{figure}
  \centerline{ \includegraphics[width=9cm]{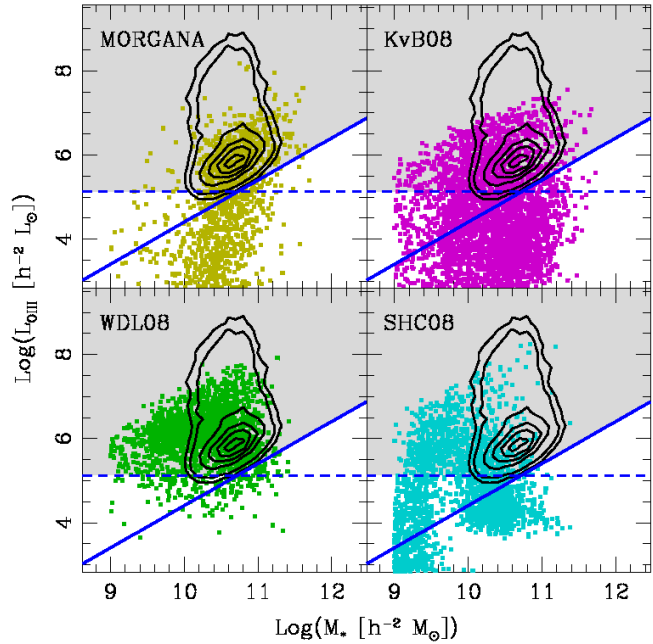} }
  \caption{Selection criteria for the AGN activity class. Black solid
    contours refers to the $5\%$, $10\%$, $30\%$, $50\%$, $70\%$,
    $90\%$ number density levels corresponding to the observed AGNs in
    the P09 sample. Yellow, magenta, green, blue dots refer to the
    predictions of the \morgana, \kang, \munich~and \somer~models
    respectively. Blue dashed line represents the observed flux limit
    converted to intrinsic luminosity at $z=0.1$; blue solid line
    correspond to the $S/N$ cut of the data. The shaded area
    corresponds to the final selected region.}
  \label{fig:o3sele}
\end{figure}

Predicted accretion rates onto SMBHs are converted into an estimate
for the luminosity of the \oiii $5007 \AA$ line ($L_{\rm \oiii}$) as
illustrated below. For each model galaxy, we compute the expected
bolometric luminosity associated with each accretion event in the
``bright-mode'' as:
\begin{equation}
L_{\rm bolo} = \epsilon \dot{m}_{\rm QM} c^2
\end{equation}
where we assume $\epsilon = 0.1$. We then estimate $L_{\rm \oiii}$
using an empirical approach based on the results obtained by
\citet{Heckman05} from a sample of type I and II AGNs. These authors
estimate the mean luminosity ratio between the hard X-ray luminosity
($L_{\rm HX}$) and $L_{\rm \oiii}$ for a sample of X-ray selected
local AGNs, and find that:
\begin{equation}\label{eq:heckman}
 \log(L_{\rm HX} / L_{\rm \oiii})\sim 2.15
\end{equation}
To make use of this result, we first convert bolometric luminosities
into $L_{\rm HX}$ by means of the bolometric correction proposed by
\citet{Marconi04}. We also test an alternative conversion based on the
bolometric correction proposed by \citet{Wu09} for a sample of type I
AGNs:
\begin{equation}\label{eq:wu}
\log(L_{\rm bolo}) = 0.95 \log(L_{\rm \oiii}) + 5.39
\end{equation}
and we find that the two approaches provide similar results. We note
that the \citet{Heckman05} sample includes both Type I and type II
AGNs, while the \citet{Wu09} sample includes only Type I AGNs. Since
type I AGNs were explicitly excluded in the P09 analysis, we will show
only results based on the \citet{Heckman05} conversion in the
following. In addition, following \citet{Simpson05}, we estimate the
Type I fraction as a function of $L_{\rm \oiii}$ and we use it as a
statistical correction to account for the removal of Type Is in
P09. We have verified that this correction does not affect our
conclusions.

In order to compare model predictions with the P09 results, we also
need to apply similar `selection criteria'. The main requirement for a
galaxy to belong to the AGN class in the P09 analysis is the detection
of the \oiii emission line in the SDSS spectra, above a well defined
signal-to-noise level $S/N > 3$. This implies that, if the \oiii flux
is relatively weak compared to the continuum, and the galaxy is at a
relatively large distance (so that its total flux is low), the
continuum will be noisy and the emission line will not be detected at
sufficient $S/N$. In fig.~\ref{fig:o3sele}, we plot the distribution
of model galaxies (coloured symbols) in the $L_{\rm \oiii}$-stellar
mass plane, and compare it with the distribution measured from the
SDSS: black contours show the $5$, $10$, $30$, $50$, $70$, and $90$
per cent number density levels from the observed data. This figure
illustrates the consequences of assuming a $S/N$ limit: \oiii lines of
comparable luminosity are more easily detected in less massive
galaxies, due to the lower continuum level. Using the distributions
shown in fig.~\ref{fig:o3sele}, we define a composite selection
criterion for our model galaxies. We first convert the observed flux
limit ($f_{\rm \oiii} = 5 \times 10^{-17} {\rm erg/s/cm}^2$,
de-reddened as described in \citealt{Kauffmann03b}) into an intrinsic
luminosity limit at $z=0.1$. This limit is marked with a blue dashed
line in fig.~\ref{fig:o3sele}. We then define a cut in specific
luminosity, $L_{\rm \oiii} / M_\star > 10^{5.6} L_\odot/M_\odot$,
which mimics the $S/N$ selection (blue solid line) and is chosen to
approximately follow the $5$ per cent number density level of the
observed distribution. The shaded area in fig.~\ref{fig:o3sele}
highlights the region occupied by selected model galaxies.

We note that predictions from the four models used in this study
populate different regions of the $L_{\rm \oiii}$-stellar mass
diagram. This suggests that, in principle, we could use these
distributions to discriminate between competing models. Unfortunately,
however, observational data do not cover a region of this space that
is wide enough to provide conclusive evidence in favour of or against
a given model. As an example, \munich~and \kang~show similar results
in the high luminosity region: this is expected, since these two
models use very similar prescriptions for gas accretion during merger
events. In \morgana, the distribution of AGNs shows a strong
dependence on $M_\star$, similar to that obtained for radio galaxies
(see sec.~\ref{sec:radio}). We remind that in this model gas accretion
is treated in the same way in both regimes, the only difference being
the amount of accreted gas. The correlation between $L_{\rm \oiii}$
and $M_\star$ originates from the strong dependence of gas accretion
on $M_{\rm BH}$ and $\sigma_B$.

\subsection{Star formation rate and $\ha$ line strength}\label{lha}
\begin{figure}
  \centerline{ 
    \includegraphics[width=9cm]{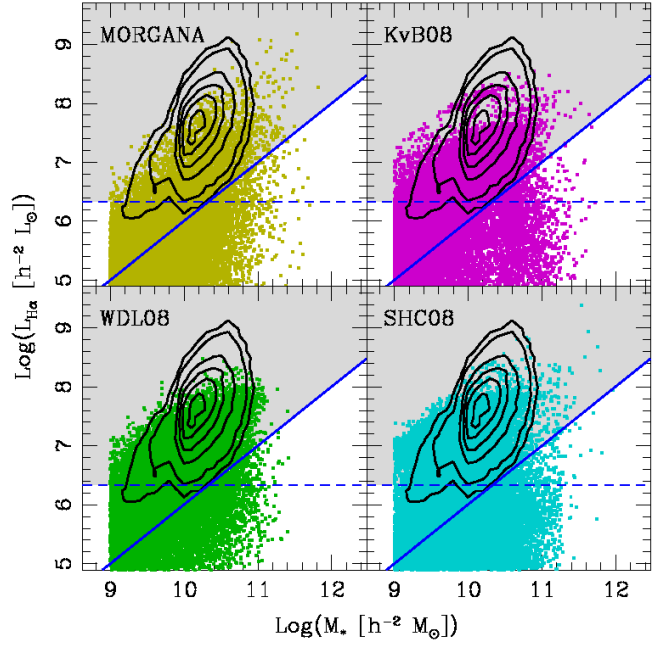} 
  }
  \caption{Selection criteria for the SFG activity class. Symbols,
    lines and shading as in fig.~\ref{fig:o3sele}}
  \label{fig:hasele}
\end{figure}

Predicted star formation rates are converted into $\ha$ line
luminosities ($L_{\rm \ha}$) using the observationally calibrated
relation proposed by \citet{Kennicutt98}:
\begin{equation}
SFR (M_\odot / {\rm yr}) = 7.9 \times 10^{-42} L_{\rm \ha} ({\rm erg / s})
\end{equation}
Since the $\ha$ luminosity provides, on average, an estimate of only
half the total SFR, assuming an average $50\%$ escape fraction of
ionising photons \citep{Kennicutt98}, we reduce the $L_{\rm \ha}$
estimated using the above equation by a factor of two. In
fig.~\ref{fig:hasele}, we show the distributions of predicted $L_{\rm
  \ha}$ luminosities, and compare them with the observed
distribution. In order to compare model predictions with the data
presented in P09, we adopt a strategy similar to that illustrated in
sec.~\ref{lo3}. We first convert the apparent flux limit ($f_{\rm \ha}
= 8 \times 10^{-16} erg/s/cm^2$ de-reddened as described in
\citealt{Kauffmann03b}), into an intrinsic luminosity at $z=0.1$ (blue
dashed line). We then define a cut in specific luminosity $L_{\rm \ha}
/ M_\star > 10^{-3.9} L_\odot/M_\odot$ (blue solid line), that is
chosen to approximately follow the $5$ per cent density level and
mimics the $S/N>3$ cut adopted in the observational analysis. As in
fig.~\ref{fig:o3sele}, the shaded region highlights the selected
area. We do not attempt to take into account other possible biases due
to fibre size (see e.g. \citealt{Brinchmann04} for a discussion): we
remind the reader that the observational estimates are based on fibre
spectra, while SAMs provide total integrated SFRs for model
galaxies. Predictions from the four models used in this study populate
very similar regions of the $L_{\rm \ha}$ - $M_\star$ diagram.

\subsection{Radio}\label{sec:radio}
\begin{figure}
  \centerline{ \includegraphics[width=9cm]{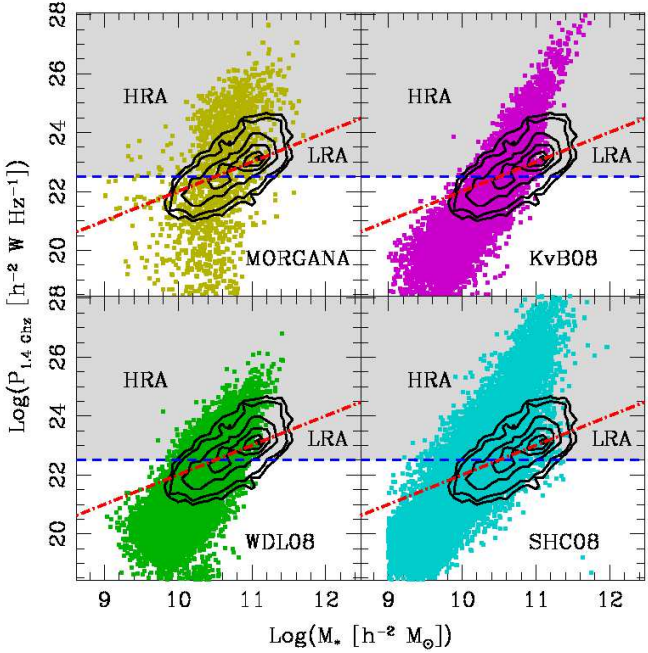} }
  \caption{Selection criteria for the Radio activity classes. Symbols,
    lines and shading as in fig.~\ref{fig:o3sele}. Red dot-dashed line
    represents the assumed threshold between LRA and HRA classes.}
  \label{fig:radiosele}
\end{figure}

In order to estimate the radio emission associated with the
``radio-mode'', we assume that an energy corresponding to $L_{\rm RM}$
is converted into mechanical work that radio AGNs exert on the
surrounding medium. We then use the relation calibrated by
\citet{Willott99} to determine the luminosity at $151 Mhz$ ($L_{151}$,
in units of $10^{28} {\rm W Hz^{-1} sr^{-1}}$):
\begin{equation}\label{eq:willott}
L_{\rm kin} = 3 \times 10^{45} f^{3/2} L_{151}^{6/7} {\rm erg/s}
\end{equation} 
In eq.~\ref{eq:willott}, $f$ is a correction factor that takes into
account the systematics in the observational measurements, and we
assume the average value $f = 15$ (\citealt{Hardcastle07}, see also
\citealt{Shankar08}).  We then compute the $1.4$ Ghz luminosity
($L_{\rm 1.4 Ghz}$) assuming a fixed spectral shape $f_{\nu} \propto
\nu^{0.7}$, and apply the same detection limits of P09 ($L_{\rm 1.4
  Ghz}>10^{21} {\rm h^{-2} W Hz^{-1}}$). The full conversion between
$L_{\rm RM}$ and $L_{\rm 1.4}$ can be thus expressed by the following
expression:
\begin{eqnarray}\label{eq:radio_tot}
L_{\rm 1.4} &=& 4 \pi \times 10^{28} \left ( \frac{L_{\rm RM}}{3 \times 10^{45} 
  f^{3/2} } \right )^{7/6} \times \nonumber \\
&\times& 10^7 \left ( \frac{1.4}{0.151}\right )^{0.7} W Hz^{-1} 
\end{eqnarray}
We have also tested an alternative method based on the empirical
conversion calibrated on the results by \citet{Best06}:
\begin{equation}\label{eq:best}
\frac{L_{\rm kin}}{10^{36} {\rm W}} = 3 \times \left ( \frac{L_{1.4
    {\rm GHz}}}{10^{25} {\rm W Hz^{-1}}} \right )^{0.4}
\end{equation}
and verified that this does not affect our conclusions.

As in P09, we split radio galaxies into two samples with low and high
radio activity (LRA and HRA respectively) using a specific luminosity
threshold of $L_{\rm 1.4 Ghz}/M_\star = 10^{12} {\rm h^{-2} W Hz^{-1}}
L_\odot/M_\odot$. We stress that in the following we do not consider
radio emission associated with high accretion rates in the models,
i.e. we neglect the radio loud fraction of QSOs (see
e.g. \citealt{Jiang07}). Our estimates of the fraction of radio
galaxies are therefore lower limits: however we want to test the
conservative hypothesis that all radio activity is linked to the
``radio-mode'' feedback. In fig.~\ref{fig:radiosele}, we compare model
predictions with the observed radio power distribution. In order to
mimic the observational selection criteria, we apply the same apparent
radio flux limit of $2.1 {\rm mJy}$. In fig.~\ref{fig:radiosele}, we
show the corresponding value for the intrinsic power at $z=0.1$, as a
blue dashed line, while the red dot-dashed line marks the threshold
between the LRA and HRA classes. The selection criteria illustrated
above influence our results, particularly for the LRA sources, but do
not affect our conclusions. We will come back to this issue later. We
note that models predict the correct dependence of $P_{\rm 1.4 Ghz}$
on $M_\star$, with more massive galaxies hosting brighter sources, but
the correlation between radio power and stellar mass appears to be
stronger in the models with respect to the observational results by
P09. The strong dependence of $P_{\rm 1.4 Ghz}$ on $M_\star$ in the
\munich~and \kang~models results from the strong correlation assumed
between the strength of ``radio-mode'' feedback and the mass of the
parent DM halo. Since this mode is active only for central galaxies,
and the stellar mass of these is strongly related to the parent halo
dark matter mass, this results in a strong correlation between the
intensity of radio-power and the galaxy stellar mass. Although
\somer~adopt a different model governing the efficiency of radio mode
activity, as already shown in \somer~their model produces a nearly
identical dependence on halo mass as that used in \munich. Therefore
we expect a similar dependence on stellar mass as well. Among the
models used in this study, \morgana~shows the weakest dependence of
$P_{\rm 1.4 Ghz}$ on $M_\star$. This is due to the fact that in this
model, gas accretion onto SMBHs in the ``radio-mode'' regime is not
directly linked to $M_{\rm DM}$, but is mainly regulated through
$M_{\rm BH}$ and $\dot{m}_{\star,B}$ (eq.~\ref{eq:morgana}).

\section{Results}\label{sec:results}
\begin{figure*}
  \centerline{ 
    \includegraphics[width=9cm]{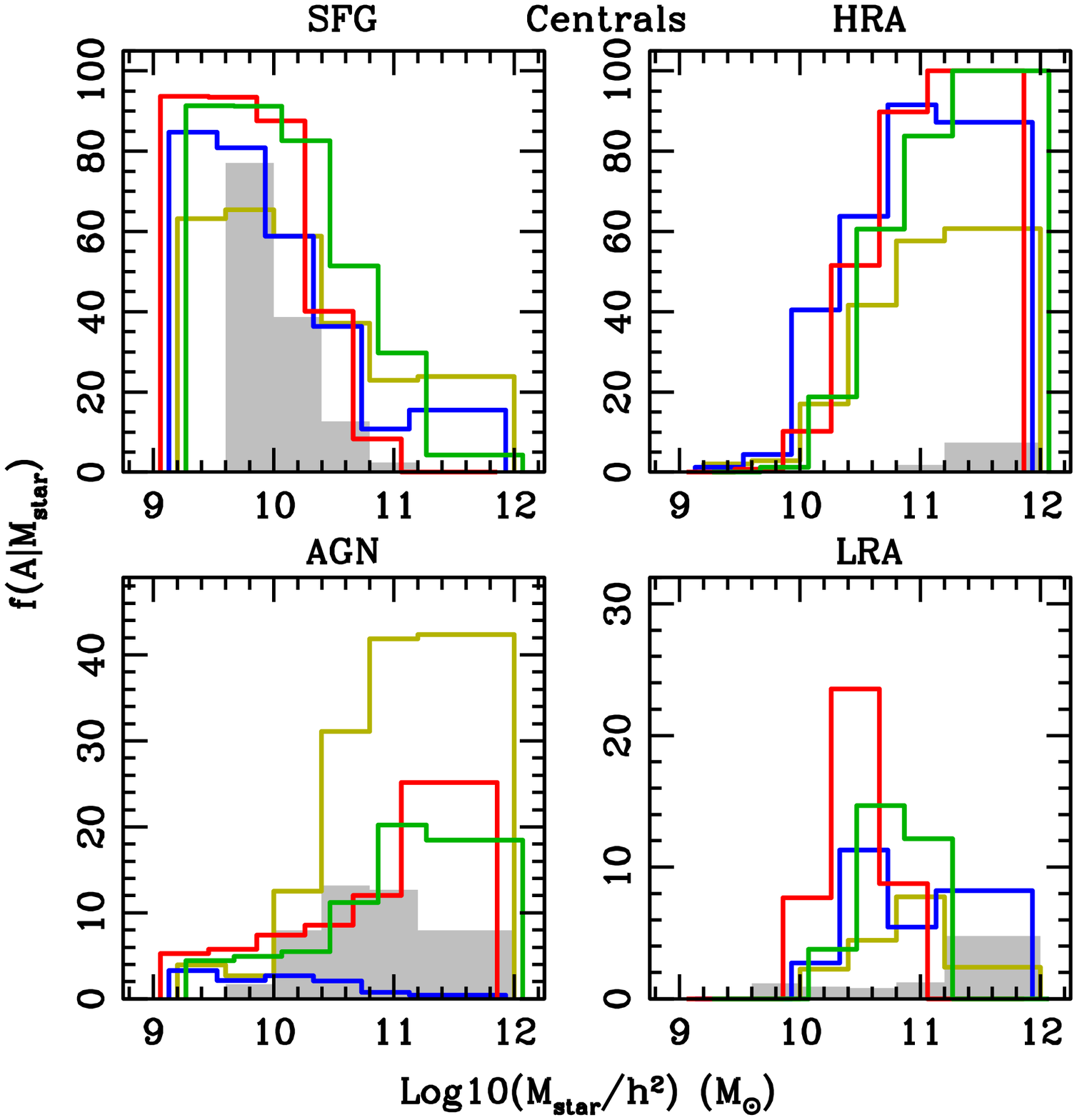} 
    \includegraphics[width=9cm]{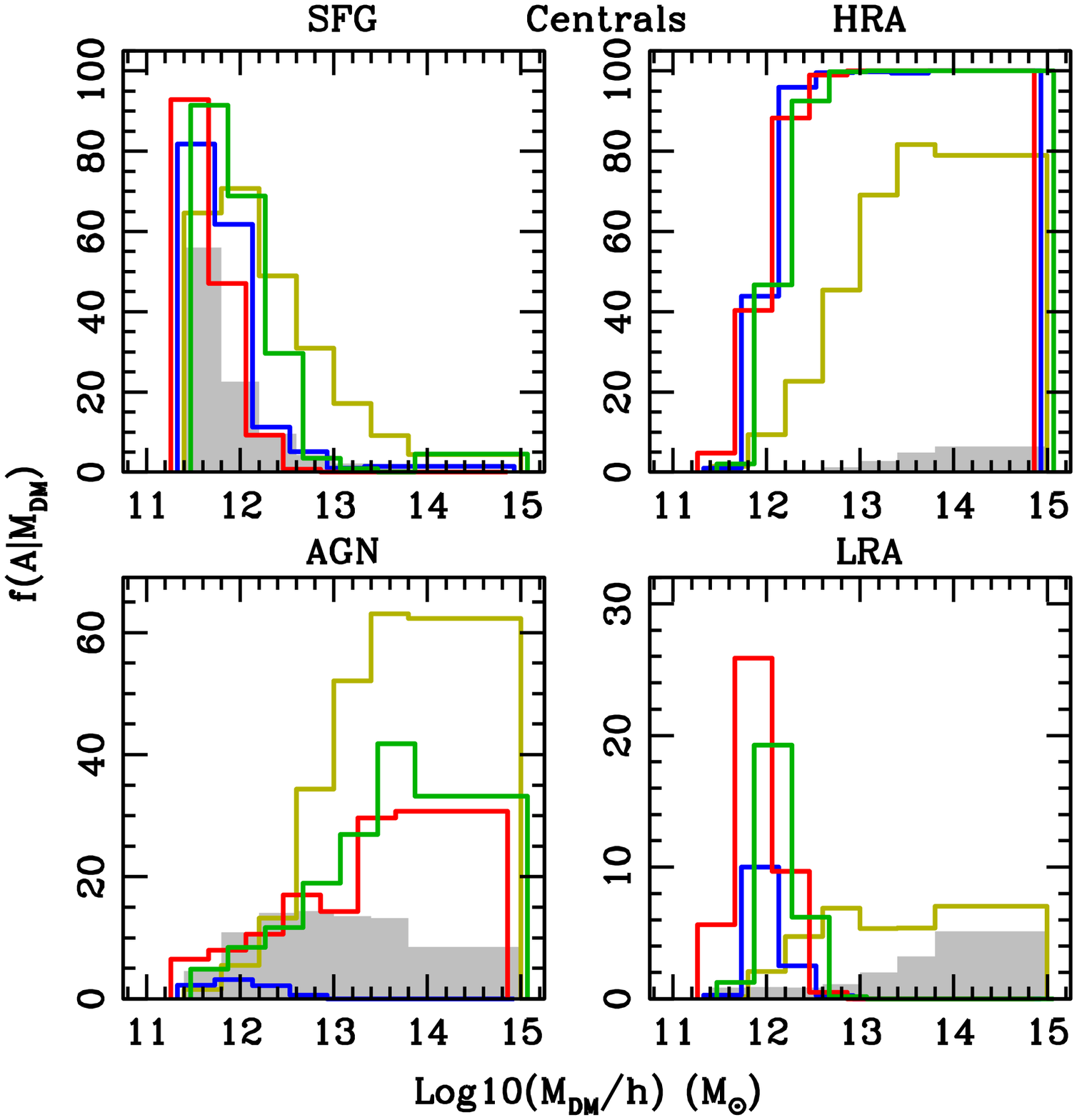} 
  }
  \caption{Fraction of central galaxies belonging to the different
    activity classes as a function of stellar mass ({\it right panel})
    and as a function of the mass of the parent DM halo ({\it left
      panel}). In all panels the yellow, red, green and blue
    histograms refer respectively to the \morgana, \kang, \munich~and
    \somer~models (histograms are slightly offset for clarity). Shaded
    areas refer to the fractions observed by P09.}
  \label{fig:frac_cen}
\end{figure*}
\begin{figure*}
  \centerline{ 
    \includegraphics[width=9cm]{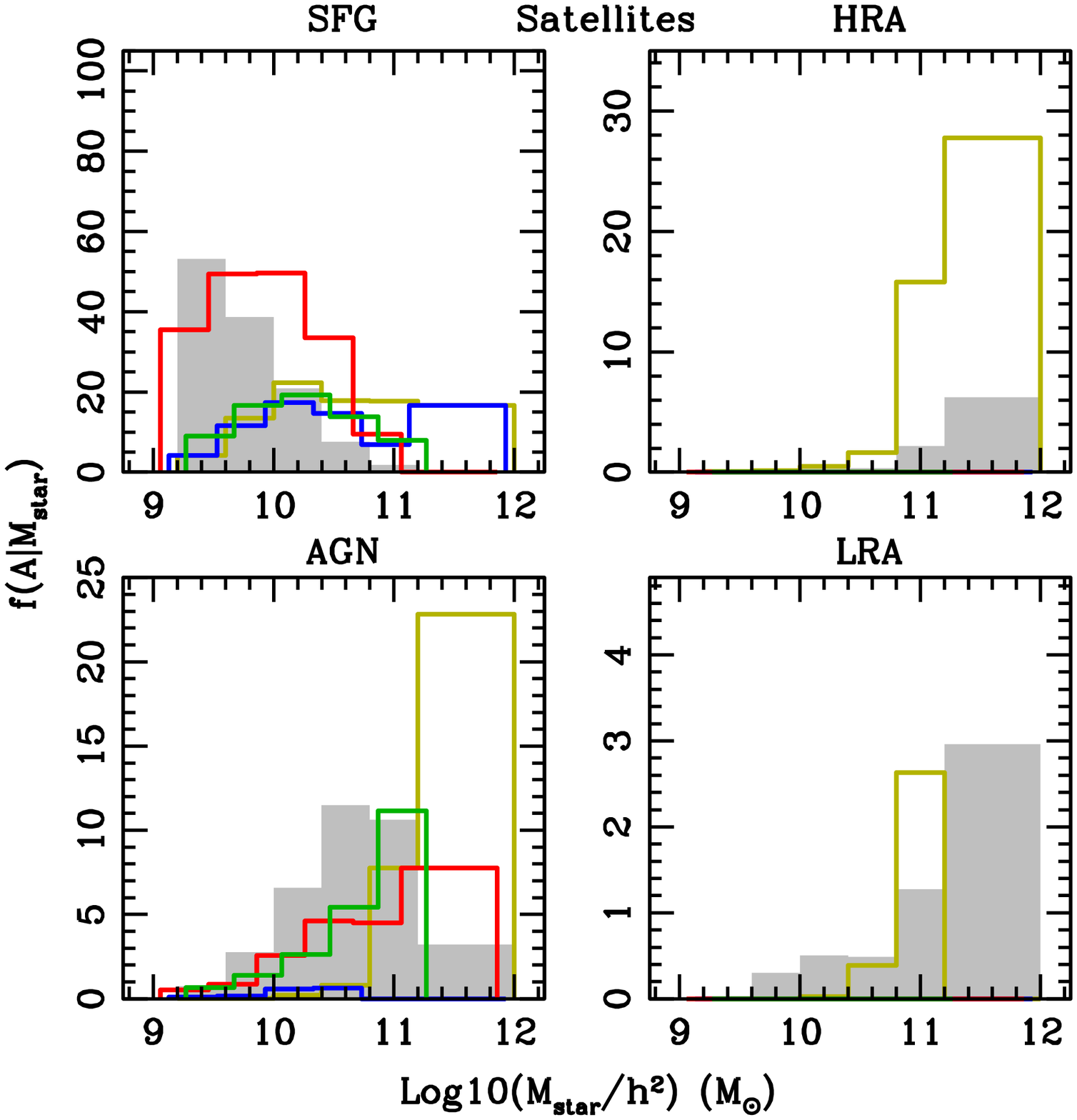} 
    \includegraphics[width=9cm]{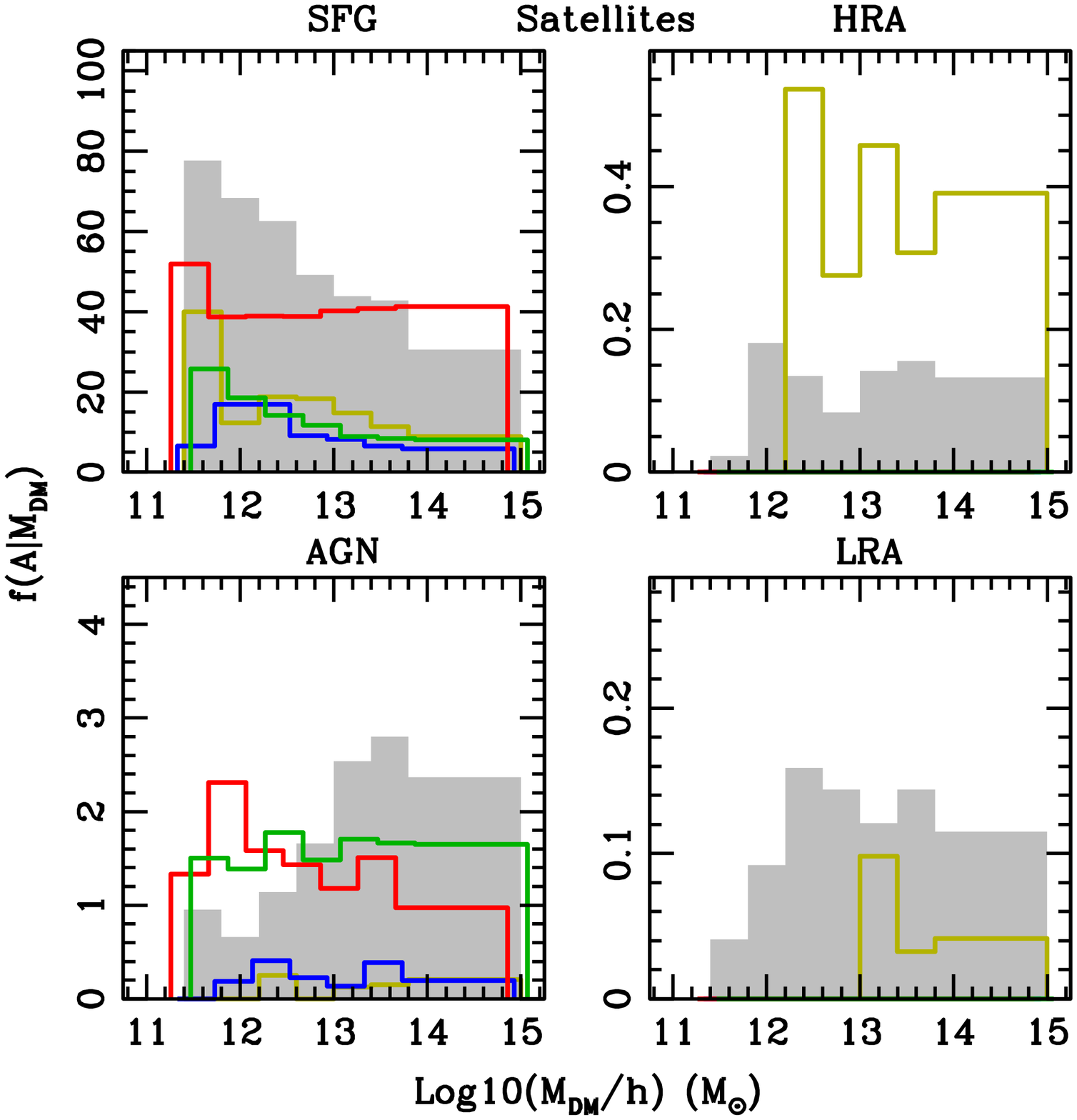} 
  }
  \caption{Fraction of satellite galaxies belonging to the different
    activity classes as a function of stellar mass ({\it right panel})
    and as a function of the mass of the parent DM halo ({\it left
      panel}). Lines and shading as in fig.~\ref{fig:frac_cen}.}
  \label{fig:frac_sat}
\end{figure*}

Following P09, we consider 4 different activity classes: SFG, AGN, LRA
and HRA (see previous sections), and consider separately central and
satellite galaxies. For each class, we define the fraction $f(A|P)$ of
model galaxies belonging to the activity class $A$ and with a given
property $P$ (i.e. stellar mass and parent halo mass). We note that in
the P09 analysis, SFGs and AGNs are separated using the BPT
diagram. In this paper, we simply assume that all model galaxies with
a detectable \oiii line belong to the AGN class. Therefore, our model
predictions for $f(AGN|P)$ represent an upper limit, since AGNs hosted
in SFGs may be misclassified in the observations. However, as we will
see in the following, we do not expect this contamination to affect
our conclusions. We note that, although the SFG and AGN classes are
mutually exclusive (as in the real data), the same objects might be
classified as radio sources, both for model predictions and for
observations.

In fig.~\ref{fig:frac_cen} and ~\ref{fig:frac_sat}, we show the
$f(A|M_{\rm DM})$ and $f(A|M_\star)$ fractions, for central and
satellite galaxies respectively. Model predictions are shown using
coloured histograms (yellow, red, green, blue for the \morgana, \kang,
\munich, \somer~models respectively), while shaded histograms show the
observational measurements by P09. When constructing the model
histograms, we apply additional cuts to our model samples by
considering only galaxies with predicted stellar masses between
$10^{9.2} < M_\star h^{-2} / M_\odot < 10^{12}$ and living in DM
haloes of $10^{11.4} < M_{\rm DM} h^{-1} / M_\odot < 10^{15}$, as in
P09. Typical uncertainties in observed fractions are of the order of
$5\%$ for the SFG and AGN classes, $10\%$ for HRA/LRA centrals, $20\%$
in $f(A|M_{\rm DM})$ for HRA/LRA satellites, and $30\%$ in
$f(A|M_\star)$ for HRA/LRA satellites.

We first consider central galaxies (fig.~\ref{fig:frac_cen}): the peak
of the mock distributions shift from low-mass haloes for SFGs to
high-mass haloes for HRA sources, in agreement with the observational
measurements. In addition, we find that the distributions as a
function of galaxy stellar mass are quite similar to those computed as
a function of halo mass. This is expected because of the strong
correlation between halo mass and stellar mass for central galaxies in
SAMs (see e.g. \citealt{Yang08}).

The figure, however, illustrates several discrepancies between model
predictions and observations. The \kang, \somer~and \munich~models
reproduce the observed decrease of the fraction of star forming
galaxies with increasing stellar and halo mass. \morgana~predicts
higher $f(SFG|M_{\rm DM})$ than observed, especially at intermediate
masses. This is due to the inefficient quenching of star formation in
massive galaxies, leading to a large fraction of active massive
central galaxies \citep{Kimm08}. Both $f(AGN|M_{\rm DM})$ and
$f(AGN|M_\star)$ are more skewed towards higher masses with respect to
the observational measurements. The \somer~model represents an
exception, with a very low predicted fraction of AGNs, and a
distribution that peaks at the low-mass end, both as a function of
stellar and parent halo mass. This is due to the modelling adopted for
QSO activity during mergers: this model considers a critical BH mass,
above which no further accretion is allowed onto SMBHs. At low
redshift, most massive galaxies already host massive BHs, so that most
mergers at low redshift happen between galaxies whose combined BH
masses are larger than the critical mass adopted in this model and
further accretion onto the remnant BHs is suppressed. In contrast, the
other three models used in this study still allow gas accretion onto
the SMBH of the remnant galaxy, as long as the progenitors still have
some gas left (i.e. if the merger is not ``dry'').

The predicted fractions of radio sources as a function of both stellar
and halo masses are larger than the observed ones by at least one
order of magnitude: all models predict that more than $80$ per cent of
central galaxies in relatively massive haloes should host a strong
radio galaxy. The disagreement with observations is particularly large
for $f(HRA|M_{\rm DM})$: almost all models predict that all $M_{\rm
  DM}>10^{13} M_\odot /h$ haloes are associated with a detectable
radio galaxy. Also the shapes of the $f(LRA|M_{\rm DM})$ and
$f(HRA|M_{\rm DM})$ distributions predicted from the models used in
this study differ from those observed (this holds also for
$f(LRA|M_\star)$ and $f(HRA|M_\star)$): bright and faint sources are
associated with larger and smaller haloes respectively, while the
observed distributions for these two classes are very similar. The
strong correlation between radio activity and halo mass is expected in
the \kang, \munich~and \somer~models, given the assumed
proportionality between accretion rates in the ``radio-mode'' and the
mass of the parent dark matter halo. It is interesting that the
typical halo mass that separates the HRA and LRA classes is similar in
these models, despite the different dependence on circular velocity
adopted. We anticipate that the precise value of this ``transition''
mass depends on the details of the conversion of heating rates into
radio luminosity (we will come back to this in sec.~\ref{sec:final}).
\morgana~is the only model that predicts a significant population of
low-luminosity radio galaxies for central galaxies in massive haloes,
but both the predicted fractions and the distributions differ from
those observed. We stress that the shape of the LRA distribution
depends on the `selection criteria' adopted for model galaxies (see
sec.~\ref{sec:radio}). A decrease in the assumed flux limit translates
into larger samples of faint radio sources: this affects only the low
stellar and halo mass tail of the distributions, leading to larger
$f(LRA|P)$ values. As a consequence, the peak of the model
distributions moves towards lower masses, increasing the discrepancy
with the observational measurements.
\begin{figure*}
  \centerline{ 
    \includegraphics[width=18cm]{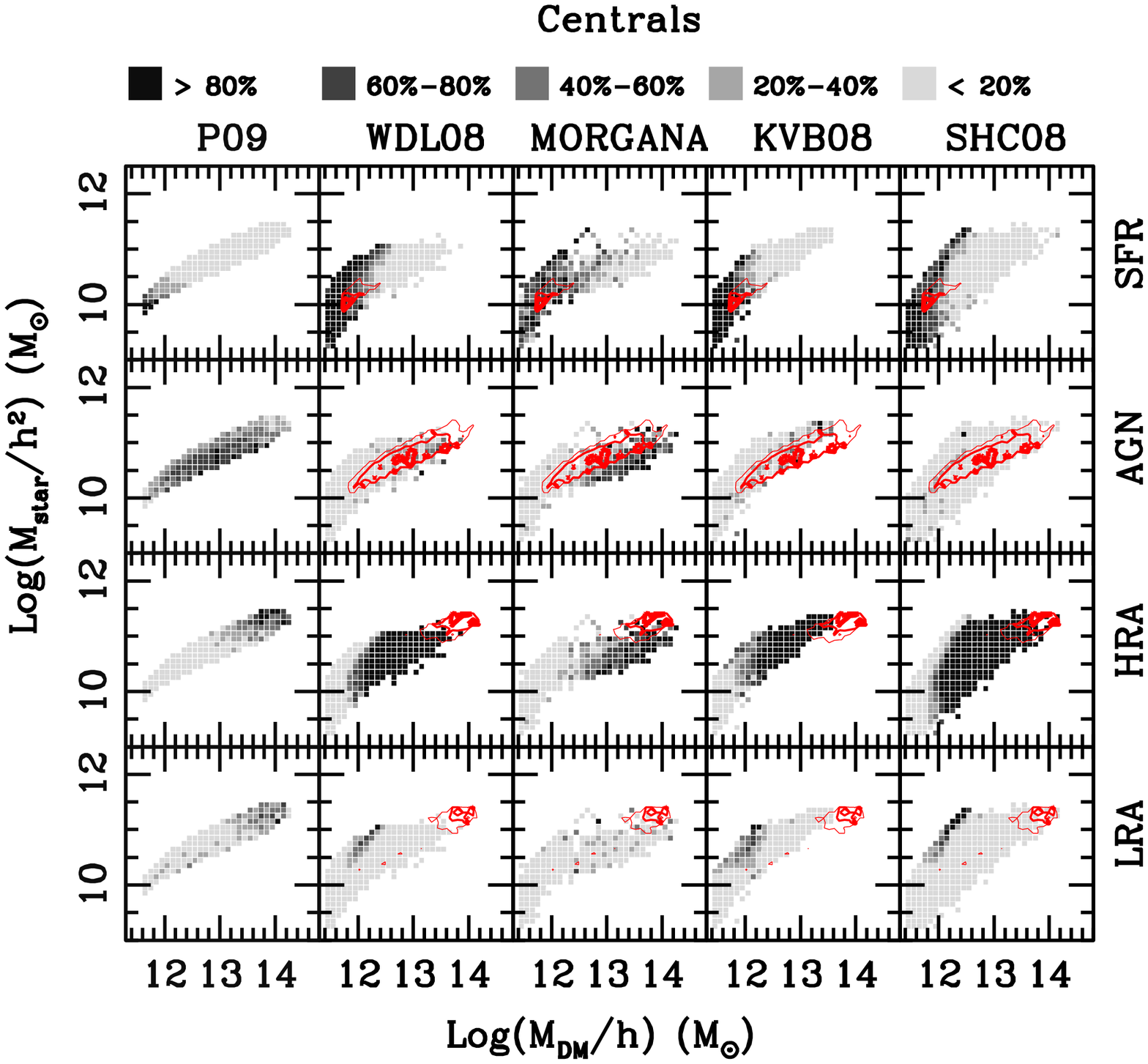} }
  \caption{2-d distribution of activity indicators for central
    galaxies on the $\log(M_\star)$ vs $\log(M_{\rm DM})$ space: gray,
    shadings refer to number density of galaxies in the P09 data,
    \citet{Wang08}, \morgana, \citet{Kang06} and \citet{Somerville08}
    models respectively. All distributions are normalized to the
    maximum value of $f(A|M_{\rm DM},M_\star)$ for each activity
    class, with darker levels corresponding to growing fractions as
    indicated in the upper legend. Red contours in each model panel
    mark the $25\%$, (thin lines) $50\%$ (solid lines) and $75\%$
    (thick lines) number density levels for the same activity class in
    P09 data.}
  \label{fig:2ddc}
\end{figure*}
\begin{figure*}
  \centerline{ 
    \includegraphics[width=18cm]{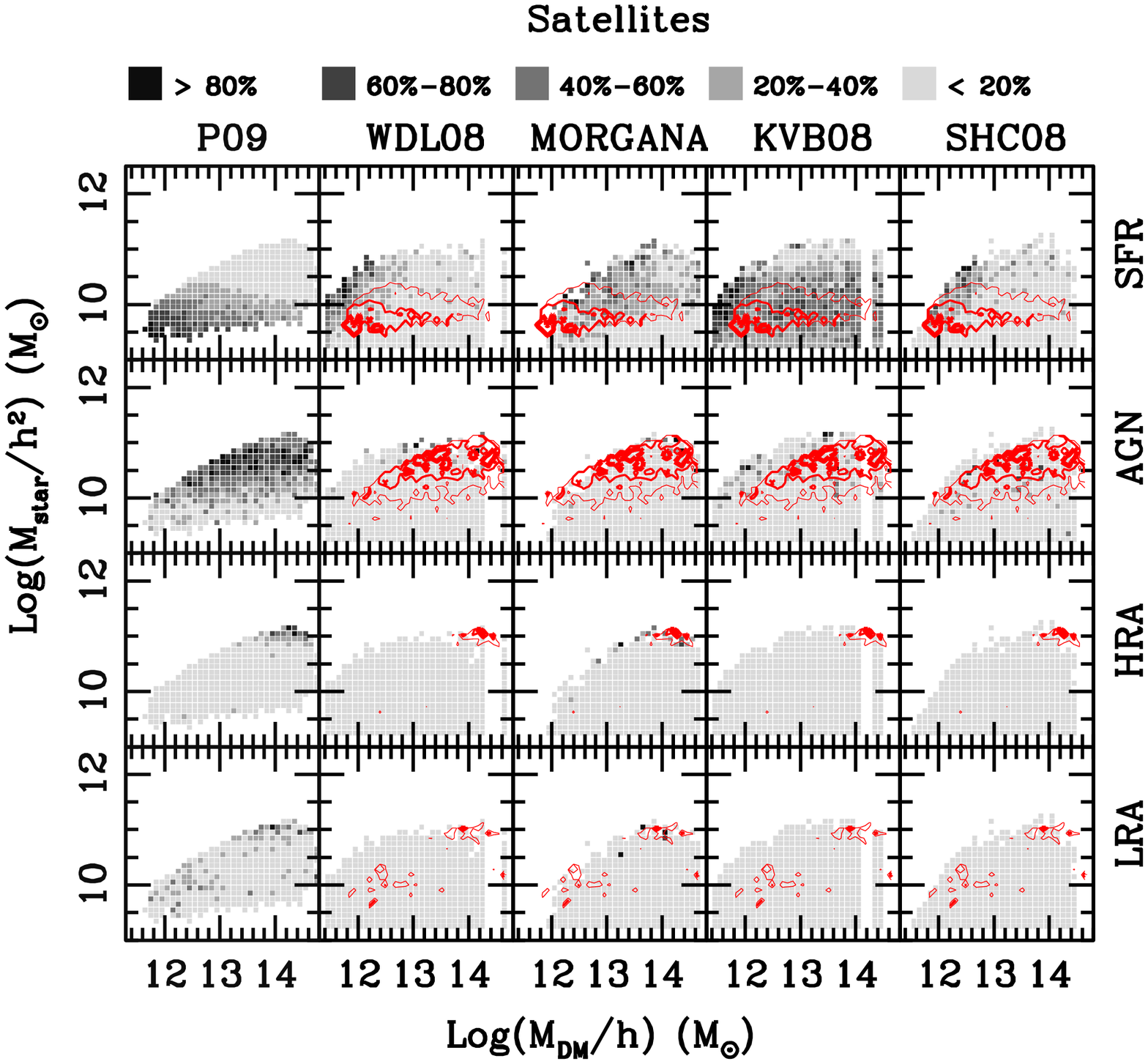} }
  \caption{Same as fig.~\ref{fig:2ddc} for satellite galaxies. Colours,
    shading contours and legend as in fig.~\ref{fig:2ddc}}
  \label{fig:2dds}
\end{figure*}

We now consider the $f(A|M_{\rm DM})$ and $f(A|M_\star)$ fractions for
satellite galaxies. Most models predict $f(SFG|M_\star)$ distributions
that differ from those observed. The distributions are significantly
affected by the assumed selection criteria at their low-mass end: when
considering all satellite galaxies above the minimum stellar mass
observed, \morgana, \somer~and \munich~models predict an approximately
constant fraction of SFGs as a function of $M_\star$. \morgana~and
\somer~predict large $f(SFG|M_\star)$ for high-mass satellites, while
the lowest fractions of SF satellites are given by the
\munich~model. Only the \kang~model predicts the correct dependence of
the fraction of star forming satellites on stellar mass, although the
predicted fractions are larger than observed. It is interesting that
the same model predicts an almost flat distribution for $f(SFG|M_{\rm
  DM})$, while the other models are able to catch, at least
qualitatively, the slight decrease of SFGs with increasing halo
mass. These results highlight that the modelling adopted in \kang,
despite being able to provide the correct fraction of satellite as a
function of galaxy colours, does not fully reproduce the observed SF
activity of satellites as a function of halo mass.

The distribution of $f(AGN|M_\star)$ is too skewed towards high
$M_\star$. The only exception is again the \somer~model, which shows a
small peak of activity at $M_\star \sim 10^{10.5} M_\odot$, but does
not reproduce the observed total fractions of satellites with AGN
activity. $f(AGN|M_{\rm DM})$ is approximately flat in the models, in
qualitative agreement with data. We stress that in the \morgana,
\munich~and \somer~models, by construction, no AGN activity is
expected in satellites, since none of them allows accretion of new
cold gas onto satellites and satellite-satellite mergers at the same
time. Therefore, the galaxy-galaxy mergers responsible for the
satellite AGNs must occur when the primary component is still the
central galaxy of its own DM halo. If the event is immediately
followed by a halo merger, and the remnant galaxy becomes a satellite,
it is classified as an AGN satellite. Thanks to its implementation of
stripping and the modeling of satellite-satellite mergers, a
population of "true" AGN satellites is present in the
\kang~model. However, as shown in fig.~\ref{fig:frac_sat}, the
contribution from this channel is quite small, i.e. $f(AGN|P)$ are
similar to those predicted by the other models.

The \kang, \munich, \somer~models do not allow ``radio-mode'' feedback
in satellite galaxies by construction. This does not hold for
\morgana~that provides a population of satellite radio galaxies with
roughly the correct distributions. In addition, in this model
$f(LRA|M_\star)$ and $f(HRA|M_\star)$ have similar shape and the
fraction of satellites with AGN activity does not vary significantly
as a function of halo mass. This is due to the assumed modelling of the
cold gas reservoir around SMBHs: recently accreted satellites, despite
being instantaneously stripped of their hot gas, are still able to
sustain a short period of accretion, until their cold gas reservoir is
completely depleted.

In fig.~\ref{fig:2ddc} and~\ref{fig:2dds} we show the bidimensional
distributions $f(A|M_{\rm DM},M_\star)$ for central and satellite
galaxies respectively. These figures provide a useful summary of the
behaviors just discussed. In these figures, darker grey levels
correspond to increasing $f(A|P)$, and are normalized to their highest
value in each activity class. In the leftmost column, we show the
observed distributions (as in P09, see their fig.6). In each model
panel, we show the contours corresponding to the $25\%$, $50\%$,
$75\%$ number density levels in observed data. The strong correlation
between stellar mass and halo mass for central galaxies is clearly
visible in fig.~\ref{fig:2ddc}, and it prevents us from decoupling the
role of environment from that of stellar mass, as discussed in
P09. Fig.~\ref{fig:2ddc} confirms that all models are able to
reproduce the different distributions of SF and HRA centrals in the
$M_\star - M_{\rm DM}$ space: the former class is dominated by
low-mass centrals, living in small DM haloes, while HRA galaxies are
preferentially associated with more massive systems. AGN centrals in
\kang, \munich~and \morgana~tend to populate the upper right region of
the diagram, while AGN centrals in the \somer~model are more likely
associated with intermediate mass systems. Finally, LRA and HRA
centrals populate different regions of this space in the models, in
contrast with observational measurements. For all models, the overall
agreement with the observed distributions is rather poor.

For satellite galaxies, the activity seems to be more strongly related
to stellar mass than halo mass, in qualitative agreement with
observational data. In particular, fig.~\ref{fig:2dds} highlights that
active satellites are preferentially found in regions of the $M_\star
- M_{\rm DM}$ diagram which are contiguous to those populated by
active centrals belonging to the same activity class. This confirms
that, in all models, active satellites are recently accreted
objects. A different treatment for hot gas stripping in newly accreted
satellites, as proposed in \kang, leads to a better description of the
distributions of SF satellites, which in this model populate a wider
area in this diagram and show a stronger dependence on $M_\star$ than
on $M_{\rm DM}$.

\section{Summary and Discussion}
\label{sec:final}

In this paper, we compare predictions from four different
semi-analytic models for galaxy formation with the observational data
discussed in \citet{Pasquali09}. These data allow, for the first time,
a careful comparison with theoretical models, also in terms of the
galaxy hierarchy (i.e. their nature of centrals or satellites). This
is important for a better understanding of the limitations and
successes of current models, in which central and satellite galaxies
correspond to rather distinct evolutionary paths. The theoretical
models considered in this study adopt different assumptions for the
various modes of AGN activity: all of them distinguish between a
bright phase (usually referred to as ``bright-mode'' or ``QSO-mode'')
and a low accretion efficiency phase which is usually associated with
the development of radio jets able to offset cooling flows (the so
called ``radio-mode''). In this paper, we focus on four activity
classes, defined by the star formation rate, optical AGN activity, and
radio emission associated with the radio-mode. Model accretion rates
onto SMBHs are converted into `observables' by means of empirical
relations calibrated on local samples. We have attempted to reproduce
the observational selections and checked the influence of different
selection criteria: the absolute fractions of galaxies in each
activity class as a function of both stellar and parent halo mass
depend on the details of both the selection criteria and the adopted
conversions. The shapes of the predicted distributions, however, are
not significantly affected.

We show that the four models used in this study predict in a few cases
quite different distributions as a function of stellar and parent halo
mass. This is interesting and demonstrates that the observational
measurements can discriminate between different physical models. In
particular, we have shown that in all models SFGs are typically
low-mass galaxies residing in low-mass haloes, while powerful radio
galaxies are expected to reside at the center of relatively massive
haloes. Although this is in qualitative agreement with observations,
we highlight a number of model ``failures'':

\begin{itemize}
\item{almost all massive centrals are predicted to host a detectable radio
  source, at variance with observations.}
\item{The $f(LRA|P)$ and $f(HRA|P)$ distributions differ from those
  observed for all models. In particular, it is possible to identify a
  typical halo mass separating a `low' from `high' radio activity (LRA
  and HRA classes). The precise value of this `transition' halo mass
  depends on the adopted modelling for AGN feedback. We stress that in
  the \munich~and \kang~models this result is not surprising, since
  accretion rates onto BHs during radio-mode are explicitly related to
  the mass of the parent halo (see eq.~\ref{eq:c06} and ~\ref{eq:k08}
  respectively).}
\item{The distributions of $f(AGN|P)$ are generally too skewed towards
  high stellar and halo masses with respect to the observational
  measurements. This results in activity levels that are too high for
  massive central galaxies in clusters.  The \somer~model is an
  exception as the predicted distribution in this case is skewed
  towards low-mass galaxies and low-mass haloes. In addition, this
  models predicts total fractions of galaxies with AGN activity that
  are significantly lower than those predicted by the other three
  models used in this study. We ascribe these differences to the
  inclusion of a critical mass for BH accretion: since at low redshift
  most massive galaxies have central SMBHs with masses already larger
  than their critical mass, further accretion onto these objects is
  suppressed.}
\item{The distribution of star forming centrals in the $M_\star -
  M_{\rm DM}$ plane is well reproduced by the \munich, \kang~and
  \somer~models, while \morgana~predicts a wider distribution of SF
  galaxies, extending towards larger stellar and halo masses. This is
  due to an inefficient quenching of star formation in massive haloes.}
\item{The \kang~model exhibits the best agreement with the observed
  distributions of star forming galaxies. In particular, the model
  predicts the correct dependence of the $f(SFG|M_\star)$ on $M_\star$
  (although with a slightly higher normalization), both for centrals
  and satellites. As the other models used in this study, however, it
  does not reproduce the observed distribution of SF satellites as a
  function of parent halo mass.}
\end{itemize}

The disagreement between the predicted and observed fractions of radio
sources represents a potentially serious problem for current models of
galaxy formation and evolution. The inclusion of "radio-mode" AGN
feedback in theoretical models of galaxy formation was originally
motivated by the need for a powerful energy source, that would be able
to counteract gas cooling at the centres of massive haloes. Most
models assume a strong dependence of radio-mode feedback on the parent
halo mass: this is essential in order to reproduce the observed rapid
cutoff at the bright/massive end of luminosity/mass function, and the
old stellar population ages of massive galaxies. As a consequence of
this assumption, these models predict that essentially all massive
galaxies should be associated with a bright radio source, while
observational data suggest that faint and bright radio sources are
found in similar environments in equal numbers.  If, as suggested by
our analysis, we need to assume a weaker dependence on halo mass, then
this physical process might not be capable of offsetting the cooling
flows alone. This is hinted at by the fact that the model which
predicts the lowest fraction of radio sources (\morgana) is, at the
same time, the model which provides the worst agreement with the
distribution of star forming galaxies.

Details of the adopted modelling for BH growth (e.g. the duty cycle)
might also play a role. In particular, the models used here neglect
completely one relevant aspect of AGN activity: the presence of well
defined {\it duty cycles} of radio sources (of the order of $10^7 -
10^8 yrs$ for radio-loud galaxies). In SAMs, both the ``radio-mode''
feedback and the cooling rate are quantities integrated over a well
defined time interval. The ``quenching of the cooling flow'' is
therefore assumed to happen on this time-scale. These assumptions are
supported by results from \citet{Best06}, who showed that the {\it
  time-averaged} energy output from recurrent radio-sources in
elliptical galaxies balances almost perfectly the energy losses from
the hot gas in the parent DM halo. If we assume that the radio sources
are active only for a fraction of the adopted time-step, the number of
sources detected as radio galaxies will be reduced.  At $z \sim 0.1$,
however, the typical time-steps adopted in all SAMs we consider are
shorter than or of the same order as the duty cycle of radio loud
galaxies, so that the expected correction should be small. Even for
models with integration time-scales larger than the typical radio-loud
duty cycle, it is not clear if a shorter ``radio-mode'' feedback
activity would be as efficient in quenching cooling flows. In fact,
the present approach assumes the heating term to perfectly offset the
cooling when the ``radio''-mode is ``on'', while the thermal evolution
of halo gas proceeds as if there had never been a radio source present
when the ``radio''-mode is ``off''. This implies that introducing a
duty cycle in these models would probably reduce the efficiency of
quenching cooling flows\footnote{We note that \morgana~already
  implements some of these ideas (although in a very simplified
  way). This model explicitly follows the build up of a gas reservoir
  around SMBHs as a result of cold gas infall. Gas is then accreted
  onto the central object on a time-scale determined by the viscosity
  of the accretion disc, and originates collimated jets, that heat the
  halo gas. The time-interval between the onset of the cooling flow
  and its quenching through jet heating could be considered the
  equivalent of a `duty cycle'. We have shown that this model performs
  slightly better in terms of the radio active fraction (not all
  massive galaxies host a bright radio source), but it is not able to
  quench cooling flows as efficiently as needed to reproduce the
  observed colours and star formation rates of massive galaxies (see
  e.g. \citealt{Fontanot07b}, \citealt{Kimm08}).}. In addition, we
note that even fixing the total fractions of radio galaxies, we would
still have the problem that their distributions do not reproduce
observational measurements. Therefore, more realistic scenarios have
to be considered. These should take into account, for example, that
radio jets may add entropy to the gas, and that after a complete radio
cycle, the halo gas would be hotter and less dense than before the
onset of the ``radio''-mode. Cooling will thus proceed at a slower
rate. Our models also do not take into account any possible ``excess''
in the radiated energy (with respect to the energy needed to quench
cooling flows), which could unbind the gas from the halo, again
suppressing cooling on a longer timescale (see e.g.,
\citealt{Bower08}). Therefore, an accurate modelling of AGN duty
cycles requires a deep revision of AGN feedback schemes in SAMs that
goes beyond the aims of this paper.

Our results suggest that either 1) more sophisticated modelling of the
triggering of radio activity and its impact on the thermodynamics of
the surrounding gas is needed or 2) other processes, such as
gravitational heating by infalling satellites or gas clumps
\citep{KhochfarOstriker08,DekelBirnboim08}, are important in
preventing the gas in massive halos from cooling.

In all the models, active satellites represent the tail of the
corresponding distributions computed for central galaxies. This is due
to our simplified treatment of satellite evolution. In particular,
three out of the four models used here assume that the hot halo
associated with a galaxy is instantaneously stripped once the galaxy
becomes a satellite. The cold gas available is rapidly turned into
stars and ejected in the surrounding medium, never being reicorporated
onto the same galaxy. The \kang~model represents an improvement, as it
allows satellite galaxies to keep their hot gas reservoir for some
fixed time after accretion.  We have shown, however, that this model
does not reproduce the correct dependence of SF activity as a function
of halo mass.

Our work confirms that statistical studies aimed at disentangling the
role of stellar mass and parent halo mass (i.e.  the ``nature'' versus
``nurture'') are hampered by the pivotal role played by central
galaxies in the models. An improved description of the evolution of
satellite galaxies is of fundamental importance to understand the
relative importance of various physical processes responsible for
galaxy activity in different environments.
\begin{figure}
  \centerline{ \includegraphics[width=9cm]{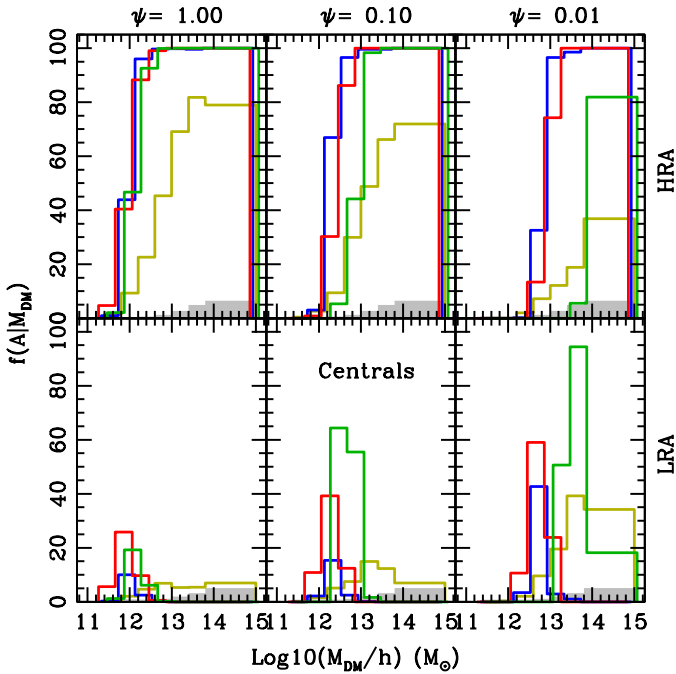} }
  \caption{Fraction of central galaxies belonging to the activity
    classes HRA and LRA as a function of the mass of the parent DM
    halo, for different choices of $\psi = L_{\rm kin}/L_{\rm RM}$. In
    all panels lines and shading as in fig.~\ref{fig:frac_cen}.}
  \label{fig:psi}
\end{figure}

A number of additional caveats should be considered. In order to
compare model predictions with observational data, we had to apply a
number of corrections and empirical conversions. In particular the
adopted conversion from estimated accretion rates into radio
luminosities is critical. To test the robustness of our results, we
repeated our analysis by assuming that only a fraction $\psi$ of the
heating energy is sufficient to originate jets and bubbles
(parametrized as $\psi = L_{\rm kin}/L_{\rm RM}$). With respect to the
results presented in fig.~\ref{fig:radiosele}, this implies lower
radio luminosities associated with the ``radio-mode'', but preserves
the shape of the predicted distributions as a function of $M_{\rm
  DM}$. We show in fig.~\ref{fig:psi} the $f(HRA|M_{\rm DM})$ and
$f(HRA|M_{\rm DM})$ distributions corresponding to three different
values of $\psi$. This figure shows that the number of sources in each
class depends significantly on the value assumed for $\psi$, but our
main conclusions are unchanged: the distributions for LRA and HRA
galaxies are different and there are too many HRA model galaxies with
respect to observational measurements. It is however interesting that
the peak of the LRA distribution tends to shift towards higher masses
as $\psi$ decreases, thus slightly reducing the disagreement with the
observed distributions.

Our analysis highlights new problems for current models of galaxy
formation and evolution, and in particular for currently adopted
models of AGN feedback. Although this mechanism helps to solve a
number of long-standing issues, it provides a poor match to the
observed fractions and distributions of radio galaxies. From an
observational point of view, a better and more complete sampling of
the activity levels as a function of different physical properties and
environment, would be crucial to confirm and strengthen our
conclusions. Important additional information will come from relating
galaxy activity with other physical properties. For example, recent
studies have suggested that the relation between activity and
morphological type can provide interesting constraints on the
accretion mechanisms \citep{Georgakakis09}, and on the co-evolution
between host galaxies and their central SMBHs \citep{Schawinski10}.

\section*{Acknowledgments}
We are grateful to Pierluigi Monaco and Francesco Shankar for useful
and stimulating discussions. FF acknowledges the support of an
INAF-OATs fellowship granted on 'Basic Research' funds. GDL
acknowledges financial support from the European Research Council
under the European Community's Seventh Framework Programme
(FP7/2007-2013)/ERC grant agreement n. 202781.

\bibliographystyle{mn2e}
\bibliography{fontanot}

\label{lastpage}

\end{document}